\documentclass[twocolumn,pra,aps,letterpaper,groupedaddress]{revtex4-1}

\usepackage{times,amsmath,amsfonts,amssymb,latexsym,textcomp}
\usepackage{color}
\usepackage{graphicx}
\usepackage{epstopdf}
\usepackage{multirow}
\usepackage{bbm}
\usepackage{tabularx}
\usepackage{amsthm}
\usepackage{dsfont}
\usepackage{float}
\usepackage{array}
\usepackage{placeins}
\usepackage{MnSymbol}
\usepackage[electronic]{ifsym}
\usepackage{tikz}
\usepackage{booktabs}
\usepackage[lofdepth,lotdepth]{subfig}
\usepackage{caption}
\usepackage{braket}

\usepackage[T1]{fontenc}
\usepackage[latin9]{inputenc}
\usepackage[english]{babel}

\newcommand{\new}[1]{{\color{black}{#1}}}
\newcommand{\newT}[1]{{\color{black}{#1}}}

\newtheorem{theorem}{Def}

\begin{document}
\title{Practical and efficient experimental characterization of multiqubit stabilizer states}
\author{Chiara Greganti$^1$, Marie-Christine Roehsner$^1$,  Stefanie Barz$^{1,2}$, Mordecai Waegell$^3$, Philip Walther$^1$}
\affiliation{$^1$~University of Vienna, Faculty of Physics, Austria, $^2$~Present address: University of Oxford,  Clarendon Laboratory, UK,
$^3$~Institute for Quantum Studies, Chapman University, Orange, CA, US}

\begin{abstract}
Vast developments in quantum technology have enabled the preparation of quantum states with more than a dozen entangled qubits.
The full characterization of such systems demands   distinct  constructions depending on their specific type and the purpose of their use.
Here we present a method that scales linearly with the number of qubits, for characterizing stabilizer states.
Our approach allows \new{simultaneous} extraction  
of information about the fidelity, the entanglement and the nonlocality of the state and thus is of high practical relevance.
We demonstrate the efficient applicability of our method by performing an experimental characterization of a photonic four-qubit cluster state and three- and four-qubit Greenberger-Horne-Zeilinger states.  Our scheme can be directly extended to larger-scale quantum information tasks.
\end{abstract}
\maketitle
\section{Introduction}
Multiqubit states are a basic resource for present and future generations of quantum information science experiments.
In particular, $N$-qubit stabilizer (or graph) states have well-proved utility
 for one-way quantum computation and  quantum information processing
\cite{Gottesman1997,Raussendorf2001,Verstraete2004b,Briegel2009}.  As the number of particles increases, the system and its properties become significantly more complex.
In order to manipulate and exploit such entangled systems,
 it is crucial to certify the generated states with respect to the
ideal stabilizer states.
The importance of analyzing these quantum resources has led to a variety of theoretical works \cite{Hein2004,Toth2005,Tokunaga2006,Guehne2007,Wunderlich2009, Niekamp2010}.
 Each of them shows certain features of the system, e.g., fidelity, purity, and entanglement robustness, by  using the stabilizer operators or their generators \cite{Gottesman1997}.
Here we present a  compact approach which allows us to \new{simultaneously} test 
the most important properties
of the generated graph states using a minimal number of measurements.
Our method utilizes the multiparty Greenberger-Horne-Zeilinger (GHZ) theorem~\cite{Greenberger1990} for a characterization of the quantum state by constructing a Bell-type inequality.
In this work we briefly introduce  nonclassical structures, defined as the critical identity products (IDs; discussed in detail in \cite{Waegell2013, Waegell2012}) and their practical applications for: generalized proofs of the $N$-qubit GHZ theorem, estimation of
the fidelity of a state, and detection of multi-party entanglement.
In the laboratory  we experimentally generate a four-qubit cluster state and
fully analyze it through IDs. We proceed in the same way with experimental three-qubit and four-qubit GHZ states in order to \new{illustrate}  
the general utility of IDs. \new{ We show how our method relates to other methods.}
\section{Theory}
\label{sec:theory}
\begin{figure}
\centering
\includegraphics[width=0.45\textwidth]{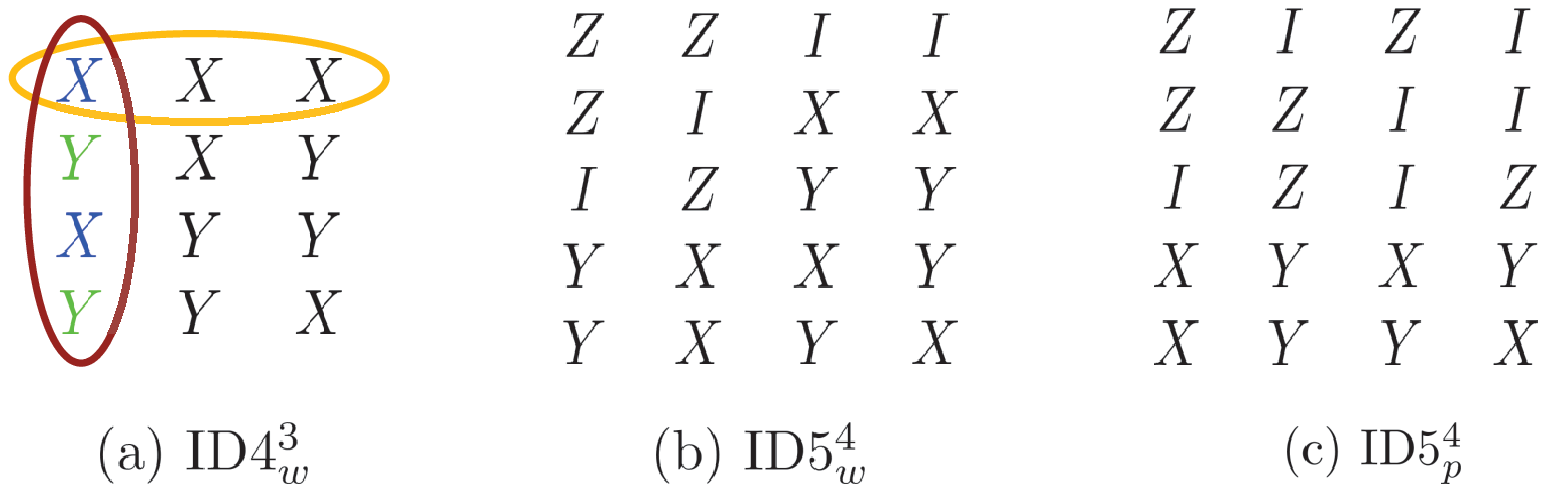}
\caption{(Color online) Critical IDs are minimal sets of mutually commuting $N$-qubit observables
that relate to specific multi-qubit states. Each row [e.g., \new{light} yellow (light gray) circle in (a)] is a different
joint observable,
 where the implied tensor product symbols are omitted for compactness,  while each column [e.g. \new{dark} red (dark gray) circle in (a)] corresponds to a different qubit. When each single-qubit Pauli observable
 appears an even number of times in each column of a negative ID, the set enables us to prove the GHZ theorem.
The tables represent (a) a whole negative ID related to the three-qubit GHZ state, (b) a whole negative ID related to the four-qubit linear cluster state, and (c) a partial positive ID related to the four-qubit GHZ state.}
\label{figure1}
\end{figure}

In the Hilbert space of $N$ qubits, nonclassical structures related to entanglement, contextuality, and nonlocality were recently introduced~\cite{Waegell2013}, which enable  addressing  foundational quantum physics topics as well as the characterization of states useful for quantum information applications.
The so-called \textit{identity products}  are the most elementary of these structures within the $N$-qubit Pauli group and
 form the constituents of the more elaborate nonclassical structures.
\begin{theorem}
IDs are sets of $M$ mutually commuting observables ($O_i$, with $i=1,...,M$) whose combined product is $\pm \mathbf{I}$ (respectively, \textbf{\textit{positive}} and \textbf{\textit{negative}} ID).
\end{theorem}
\noindent Each ID can be represented as a table $\mathbf{ID}M^N$, where each row is a different $N$-qubit observable and each \nobreak{column} corresponds to a different qubit (see Fig.~\ref{figure1}).
The rows are tensor products of single-qubit Pauli observables $o_q\equiv \{X_q,Y_q,Z_q\}$ and single-qubit identity $I_q$. When each $o_q$ appears an even number of times in all the columns, we call the full set \textbf{\textit{whole}} ID ($\mathbf{ID}M^N_w$);  otherwise, we call it \textbf{\textit{partial}} ID ($\mathbf{ID}M^N_p$).
\begin{theorem}
An ID is maximally entangled if its observables $O_i$ cannot be simultaneously tensor factorized into two or more separate IDs. It is furthermore \textbf{critical} if no deletion of \nobreak{obser}-vables and/or qubits from the set can result in a smaller ID.
\end{theorem}
\noindent This sort of entanglement is defined for a set of mutually commuting observables rather than for a particular state vector, which we can think of as the Heisenberg-picture definition of entanglement (see \newT{Appendix A}). 
As we will see, this definition of entanglement is crucial for irreducible proofs of the GHZ theorem.\\
Each ID is representative of a complete class of equivalent IDs under permutations of columns (qubits), and local transformations of qubits' coordinate systems. Every complete class of critical IDs belongs to one or more
specific classes of maximally entangled stabilizer states
\cite{Waegell2014}.  \\

\noindent \textbf{GHZ theorem}\\
Any class of ID that is whole, negative, and entangled gives a straightforward proof of the GHZ theorem for a specific class of maximally entangled $N$-qubit states and, consequently, a Bell-type inequality violation.  Following
 the $N$-qubit Mermin inequality \cite{Mermin1990}, several different approaches have been developed to study the nonlocality of multiqubit states,  particularly  graph states \cite{Guhne2005,Toth2006,Hsu2006}. In all of these works the inequality is based on stabilizer operators.
Remarkably, any whole negative entangled ID allows a
proof that is irreducible for a specific class of states and
also a generalization of the original GHZ theorem.\\
Let us consider a  joint eigenstate of a  whole negative critical ID and independent single-qubit measurements $\{X,Y,Z,I\}$ on each party. The negativity of the ID guarantees that the overall product of the expectation values of the multiqubit observables should be $-1$ according to quantum mechanics (QM).  On the other hand, the wholeness of the ID guarantees that the overall product should be $+1$ in any local hidden-variable theory (LHVT),  so we obtain the GHZ contradiction
\cite{note1}.
Figures \ref{figure1}(a) and \ref{figure1}(b) show two whole negative IDs for the three- and four-qubit cases, respectively. Note that this type of ID exists only for $N>2$ and requires measuring at most $M=N+1$ observables for a critical ID.
Starting from $N=5$, it is possible to find entangled whole negative IDs with $M<N+1$, giving the most compact demonstration of the GHZ theorem; for example,  there exist one  $\mathbf{ID}5^5_w$  and two distinct  $\mathbf{ID}5^6_w$  ~\cite{Waegell2014}.  While the original proofs of the GHZ theorem depend on the preparation of a particular state, these IDs can show the proof using any state within a particular subspace.\\

\noindent \textbf{ID Bell inequality}\\
We construct the Bell-type inequality, defining first the corresponding Bell's parameter $\alpha$ for a given negative  $\mathbf{ID}M^N_w$ as
\begin{equation}
\alpha = \sum_{i}^M \lambda_i O_i = \sum_{i|\lambda_i=1}^M  O_i - \sum_{i|\lambda_i=-1}^M O_i,
\label{alpha}
\end{equation}
where \new{ $O_i$ are the
observables of the ID and} $\lambda_i$  are the eigenvalues of a specific (target) eigenstate of the ID.
The expectation value of $\alpha$ according to QM is $\langle \alpha \rangle_{QM} = M$.
In LHVTs, the eigenvalues of each 
 $O_q$ must belong to a noncontextual value assignment, and because of wholeness
 the total number    of $O_q$ assigned to the eigenvalue $-1$ must be even.
Given this constraint, we obtain an upper bound on the expectation value of $\alpha$ in LHVTs according to
\begin{equation}
\langle \alpha \rangle_{LHVT} \leq M-2,
\label{Bell_In}
\end{equation}
which we call the \textit{ID Bell}  inequality (see \newT{Appendix A} for more details).\\

\noindent \textbf{ID entanglement witness}\\
Any Bell-type inequality can be used to experimentally verify the correlations within a multiparty state. For the two-qubit case the Bell parameter related to the Clauser-Horne-Shimony-Holt (CHSH) inequality \cite{Clauser1969} is a widely used quantity to characterize sources of two entangled qubits \cite{Kwiat1999,Altepeter2005a}.
  In a similar way the $N$-qubit ID-Bell inequality can be used to certify sources of multi-qubit entangled states.\\
  We can  construct a set of \textit{general} witness operators for each ID $\{\mathcal{W}^{ID}_\mathcal{C}\}$.  This is done by constructing $\langle \alpha \rangle$ for any particular class  $\mathcal{C}$ of states and maximizing over the entire class to obtain
$  \gamma_\mathcal{C} = \max_{|\psi\rangle \in \mathcal{C}} \langle \psi |\alpha|\psi \rangle.$
  The ID $\mathcal{C}$ witness operator is then
  \begin{equation}
 \mathcal{W}^{ID}_\mathcal{C} = \gamma_\mathcal{C}I - \alpha,
   \end{equation}
  which guarantees that $\langle\mathcal{W}^{ID}_\mathcal{C} \rangle \geq 0 $ for all states in $\mathcal{C}$, while clearly $\langle\mathcal{W}^{ID}_\mathcal{C} \rangle < 0$ only for states close to the target state (assuming $\gamma_\mathcal{C}<M$) \cite{note2}.
   This includes the so-called  entanglement witnesses \cite{Bourennane2004}, by letting $\mathcal{C}$ be the set of all biseparable states, and, more generally, the multipartite Schmidt-number witnesses \cite{Tokunaga2006}, by letting $\mathcal{C}$ include nonbiseparable states with different Schmidt numbers than the target state. For these specific classes,  we can use an existing analytic solution \cite{Bourennane2004} to put an upper bound on   $\gamma_\mathcal{C}$, $\Gamma_\mathcal{C}$, as shown in \newT{Appendix A}.
  However, using this method, we obtain a bound that is based solely on the target state,  with no advantage of considering one ID within the set of stabilizer observables over another.  
In some cases maximizing $\gamma_\mathcal{C}$ directly for a particular ID gives a stronger discrimination than using $\Gamma_\mathcal{C}$. A general  analytic method for performing this direct maximization is an open question, but numerical methods remain feasible for many cases, such as the ones presented below.  \\

 \noindent \textbf{ID fidelity estimation}\\
 The measured value of the ID Bell parameter $\langle \alpha \rangle_{exp}$
 enables us to put a lower bound on the fidelity of an experimentally prepared state $|\psi\rangle$ with respect to the intended eigenstate $|\kappa_0\rangle$.
For a general $\mathbf{ID}M^N$ (provided that it contains $M-1$ independent generators \newT{from the
stabilizer group}), we consider the case that $|\psi\rangle$ is a pure state expressed in the eigenbasis of the ID,
\begin{equation}
|\psi\rangle = a|\kappa_0\rangle + \sum_{i=1}^V b_i |\kappa_i\rangle,
\label{fid0}
\end{equation}
where $|\kappa_i\rangle$ are the $V-1$ other eigenstates in the basis,
and $|a|^2 + \sum_{i=1}^V |b_i|^2 = 1$.
Using  $\langle \alpha \rangle_{exp}$, we obtain a lower bound on the amplitude of $|\kappa_0\rangle$ and, consequently, on the fidelity of  state $|\psi\rangle$ (see \newT{Appendix A} for the derivation):
\begin{equation}
|a|^2 \geq (\langle \alpha \rangle_{exp} - M + 4)/4 \equiv F_{ID}.
\label{FidBound}
\end{equation}
This can be generalized for mixed states by 
replacing the left side of  inequality (\ref{FidBound}) with
$\langle|a|^2\rangle \equiv \sum_{j=1}^m c_j |a_j|^2$, which is the weighted average amplitude of $|\kappa_0\rangle$ among the pure states that make up the density matrix plus noise,  $\rho = c_0 I/2^N + \sum_{j=1}^m c_j |\psi_j\rangle\langle\psi_j|$, with $|\psi_j\rangle$ being  equal to (\ref{fid0}) and $\sum_{j=0}^m c_j =1$.
In practice the bound can be used to certify the preparation of a specific quantum state using only a maximum of $N+1$ measurement settings, without resorting to complete quantum state tomography (QST) \cite{James}, which requires  $3^N$ measurement settings.  \\
%

We also want to emphasize that the critical IDs are nonclassical structures by definition. Critical whole negative IDs combine all the above-mentioned quantum properties at once. But even noncritical IDs, partial IDs, and/or positive IDs can show one or more quantum aspects of the considered eigenbasis.  Specifically, any ID that contains $N$ independent generators, whether it is critical or not, gives us a lower bound on the fidelity and can also be used for entanglement discrimination.
\section{Experiment and Results}
\label{Experiment}
\begin{figure}
\centering
\includegraphics[width=0.45\textwidth]{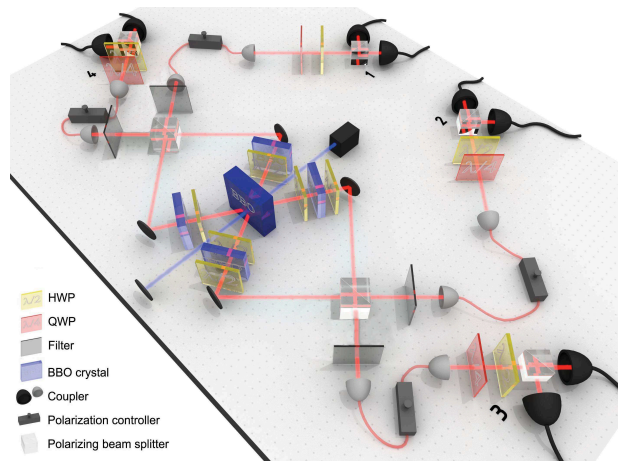}
\caption{(Color online) A femtosecond-pulsed UV-laser beam passes twice through a  $\beta$-barium borate (BBO) crystal, producing pairs of polarization-entangled photons. The photons are emitted in forward and backward directions and are recombined on polarizing beam splitters (PBSs). Walk-off effects are compensated using HWPs and half-thick BBOs. Additional HWPs  
set the entangled pairs to a selected Bell state. By postselecting fourfold coincidence events we obtain the desired cluster state or GHZ state. Polarization analysis is implemented with motorized tomographic optic components.}
\label{figure2}
\end{figure}

We apply the ID method to characterize an experimental four-qubit cluster state, related to the $\mathbf{ID}5^4_w$ [Fig.\ref{figure1}(b)], where the cluster
 state is a specific class of graph states \cite{Raussendorf2001}.
As a further demonstration of the functionality of IDs we also analyze the three- and four-qubit GHZ states, using the corresponding $\mathbf{ID}4^3_w$  [Fig.\ref{figure1}(a)] and $\mathbf{ID}5^4_p$  [Fig.\ref{figure1}(c)], respectively.
In order to generate these entangled states
 we use a photonic setup (Fig.~\ref{figure2})
 in a so-called railway-crossing configuration. Due to its compactness and high stability, this arrangement has been proven to be very suitable for several experiments \cite{Walther2005a, Prevedel2007a, Walther2005, Barz2014}.
  The scheme is based on a double spontaneous parametric down-conversion process (SPDC), bulk optics, and motorized tomographic elements to
achieve reliable measurements over long periods.
 Additional half-wave plates (HWPs) allow us to
 to switch from the generation of cluster states to GHZ states.\\

\noindent \textbf{Four-qubit linear cluster state}\\
By aligning to produce $|\phi^-\rangle$ entangled pairs in the forward direction and $|\phi^+\rangle$ in the backward direction (see Fig.~\ref{figure2} and  Ref.~\cite{Barz2014} for details), where $|\phi^\pm \rangle= (|00\rangle \pm |11\rangle)/ \sqrt{2}$,  we obtain the state:
\begin{equation}
|C_{lin} \rangle= (|0000\rangle+|0011\rangle+|1100\rangle-|1111\rangle)/2,
\label{clusterLin}
\end{equation}
which is equivalent to the linear cluster state up to local unitaries (LU), specifically up to $H\otimes I \otimes I \otimes H $, where $H=(Z+X)/\sqrt{2}$ is the Hadamard gate.
The polarizing beam splitters (PBSs) and the two interferometers in the setup, which are  necessary to select the above four terms of the state, reduce the four-fold count rate to $0.33$Hz.

\noindent \textit{Test of GHZ theorem}\\
Each of the $\mathbf{ID}5^4_w$  measurements is acquired for $4800$ s.
We obtain $\langle\alpha\rangle_{exp}=3.24 \pm 0.05$, which shows a violation of the ID Bell inequality by  $4.8\sigma$ and consequently proves the GHZ theorem for a four-qubit entangled state [Fig.\ref{figure3}(a)]. More detailed results are reported in \newT{Appendix B}. The uncertainty, like all others reported below, is due to Poissonian counting statistics and constitutes a lower limit for the errors.

\noindent  \textit{ID entanglement witness}\\
In order to certify the cluster state through ID entanglement witnesses, one constructs $\gamma_\mathcal{C}$ for any general pure quantum state.
 From the analytic method \cite{Bourennane2004} we find that to discriminate against all biseparable states ($\mathcal{B}i$), as well as the four-qubit GHZ  and W states,   $\Gamma_{\{\mathcal{B}i,GHZ,W\}}=3$ (which also coincides with $ \alpha_{LHVT}=3$), while to rule out certain other maximally entangled four-qubit states $\Gamma_{4q\mathcal{C}}=4$ \newT{\cite{note6}}. The measured value of $\langle\alpha\rangle_{exp}$ enables us to obtain a  negative value for $\langle\mathcal{W}^{ID}_{\{\mathcal{B}i,GHZ,W\}} \rangle$ but not for $\langle\mathcal{W}^{ID}_{4q\mathcal{C}} \rangle$.
 \new{In some cases we can find better (more negative) values of $\langle \mathcal{W}^{ID}\rangle$ for some specific classes of states by using numerical maximization of $\langle \psi |\alpha | \psi \rangle$ to put an upper bound on $\gamma_\mathcal{C}$. } A detailed analysis is reported in \newT{Appendix B}. In Fig.~\ref{figure3}(a) we show a few results of $\gamma_\mathcal{C}$ obtained via numerical maximization.
We consider product-states, the GHZ state $|GHZ_4\rangle$, the W state $|W_4\rangle$, and also different types of cluster states, since  the linear cluster $|C_{lin}\rangle$ is not fully symmetric under the exchange of qubits. In particular,
exchanging the order of the qubits, we evaluate $\gamma_\mathcal{C}$ for the $Z$ cluster $|C_{\text{\FallingEdge}}\rangle $  and the $shear$ cluster $|C_{\utimes}\rangle$.  The analytic method gives $\Gamma_{\{C_{\text{\FallingEdge}},C_{\utimes}\}} = 3$.\\
For four qubits there are an  infinite number of entanglement classes  that are inequivalent to one another under \newT{stochastic local
operations and classical communication (SLOCC) \cite{Dur2000a}}. All of these classes can be given in terms of a relatively small number of continuous entanglement monotones \cite{Verstraete2002}, 
 but a general classification for more qubits is not known.  A more comprehensive calculation is required to obtain the upper bound, $\gamma_\mathcal{C}$ for such states.
In any event our results for $\langle\mathcal{W}^{ID} \rangle$ certify the four-party entanglement and  rule out other particular maximally entangled four-qubit states.

\noindent  \textit{Fidelity estimation}\\
Using Eq.(~\ref{FidBound}) for the $\mathbf{ID}5^4_w$ and
$\langle\alpha\rangle_{exp}$, we estimate
$F_{ID}=0.56 \pm 0.01$.
Here we want to point out that the stabilizer group of the cluster state contains eight different $\mathbf{ID}5^4_w$'s that are equivalent by definition, and thus each of them allows for a quantum state estimation.
 All of these sets report similar values of $F_{ID}$ (see \newT{Appendix B} for the complete data).
In order to verify the validity of this bound we reconstruct the full density matrix through QST with an acquisition time of $600$s for each measurement setting.  The extracted quantum state fidelity is $F_{QST}=0.629\pm 0.007$.\\

\begin{figure}
\centering
\includegraphics[width=0.45\textwidth]{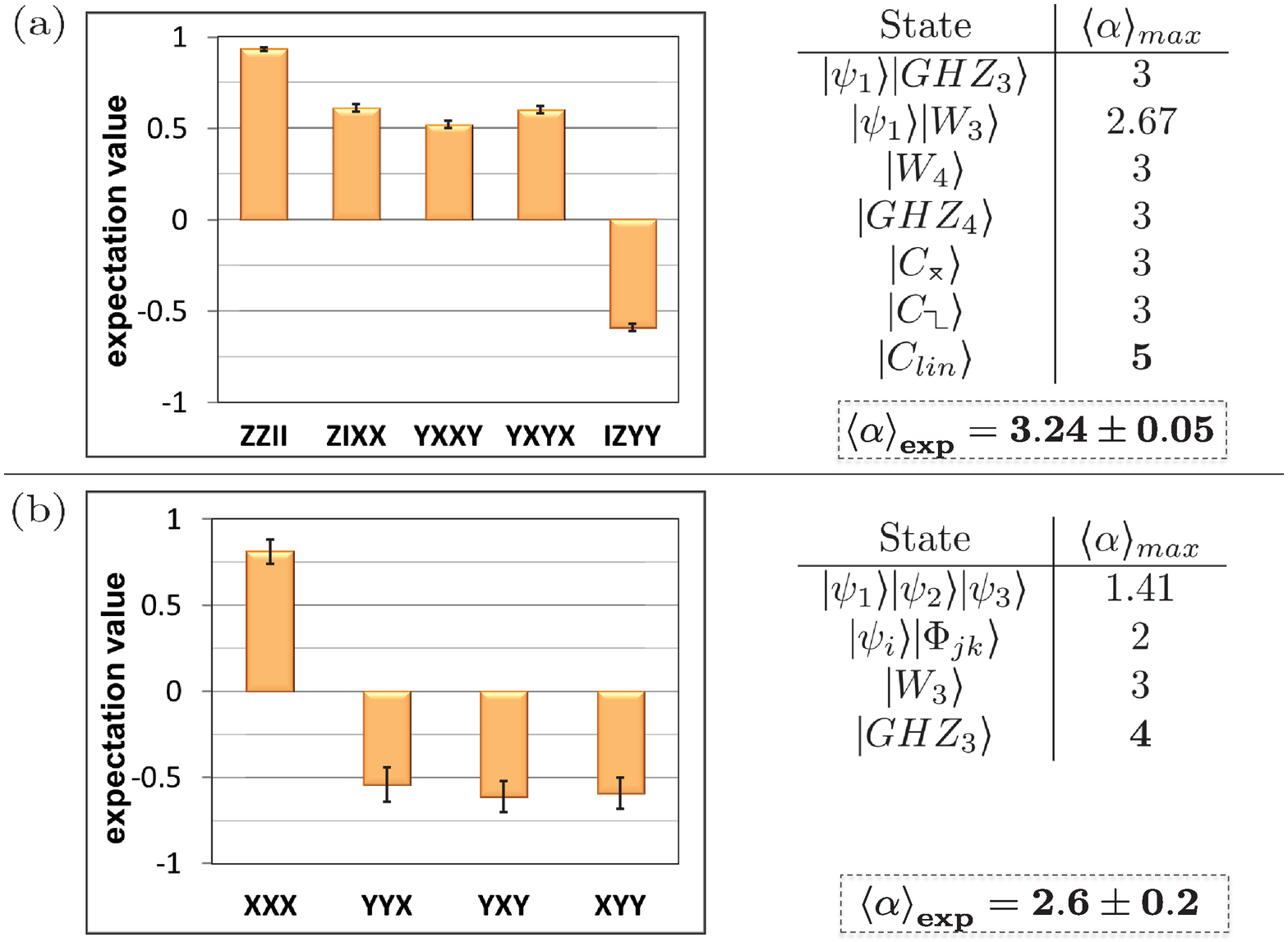}
\caption{(Color online) (a) Measured expectation values for the $\mathbf{ID}5^4_w$ (on the left)  and results of the maximization of $\gamma_\mathcal{C}$ for different four-qubit entangled states (on the right). $\scriptstyle{|C_{\utimes}\rangle=(|0000\rangle+|0101\rangle +|1010\rangle -|1111\rangle)/2}$ and $\scriptstyle{|C_{\text{\FallingEdge}}\rangle=(|0000\rangle+|0110\rangle +|1001\rangle-|1111\rangle)/2}$  are reached by exchanging the order of qubits in the linear cluster state. In the dashed box we report the experimental result of the ID-Bell parameter. (b)
 Measured expectation values for the $\mathbf{ID}4^3_w$  (on the left) and results of the maximization of $\gamma_\mathcal{C}$ for  three-qubit entangled states (on the right). In the dashed box we report the experimental result of the ID-Bell parameter.}
\label{figure3}
\end{figure}

\noindent \textbf{Three-qubit GHZ state}\\
Measuring one of the cluster state qubits and performing LU transformations, we produce the three-qubit GHZ state:
\begin{equation}
|GHZ_{3} \rangle= (|000\rangle+|111\rangle)/\sqrt{2}.
\label{3qGHZ}
\end{equation}
In the experiment we project the second qubit from Eq.(6) onto the diagonal state $|-\rangle=(|0\rangle-|1\rangle)/\sqrt{2}$ and apply a Pauli-X operation and a Hadamard gate
 on the first qubit as  postprocessing.
The state is characterized by the $\mathbf{ID}4^3_w$. We analyze it following the same procedure used for the cluster state. The GHZ theorem is proven by a violation of the ID Bell inequality of $3.1\sigma$. The ID Bell parameter is $\langle\alpha\rangle_{exp}=2.6 \pm 0.2$. We report the $\gamma_\mathcal{C}$ values for the entanglement witness in  Fig.~\ref{figure3}(b), with $\Gamma_\mathcal{C}=2$ for biseparable states. The obtained $\langle\alpha\rangle_{exp}$ is not sufficient to rule out the three-qubit $|W_3\rangle$; nevertheless, it can still confirm the three-party entanglement of the generated state.
The fidelity values obtained from the ID and QST are $F_{ID}=0.64\pm 0.05$ and $F_{QST}=0.672\pm 0.015$.
 Note the relative error for the fidelity bound is higher than that for the cluster case, since the data are determined from the tomography measurements and so are acquired in less time ($600$s).  See \newT{Appendix B} for detailed data.\\

\noindent \textbf{Four-qubit GHZ state}\\
Aligning the two entangled pairs in the setup (Fig.\ref{figure2}) to  a $|\phi^+\rangle$ state and a $|\psi^+\rangle$ state, with $|\psi^+ \rangle= (|01\rangle + |10\rangle)/ \sqrt{2}$,  the fourfold coincidences correspond to the four-qubit GHZ state up to two local unitary Pauli-X operations:
\begin{equation}
|GHZ_{4} \rangle= (|0000\rangle+|1111\rangle)/\sqrt{2}.
\label{4qGHZ}
\end{equation}
We experimentally implement these LU transformations by using HWPs for the third and fourth qubits of the state.
The state is described by the $\mathbf{ID}5^4_p$  [see  Fig.\ref{figure1}(c)], which is  critical and  partial. This implies that the IDs analysis cannot include a proof of the GHZ theorem. However, the $\mathbf{ID}5^4_p$  is still maximally entangled. It generates the complete stabilizer group of the GHZ state, so it can be exploited to test the fidelity of the state and as an entanglement witness.
We obtain a bound of the fidelity of $F_{ID}=0.71\pm 0.01$ and reconstruct the exact fidelity via QST with the result $F_{QST}=0.701\pm 0.008$.
The analytic bound for the ID witness, $\Gamma_\mathcal{C}=3$, and  $\langle \alpha\rangle_{exp}= 3.84\pm 0.05$, combine to form a negative $\langle \mathcal{W}^{ID} \rangle$ over the class of biseparable states. Additionally, the numerical maximization technique reports a maximum of $\gamma_\mathcal{C}=3$ for several maximally entangled four-qubit states (see \newT{Appendix B} for the numerical results), allowing $\langle \alpha\rangle_{exp}$ to discriminate these from the generated state. 
%
\section{Comparison of Different Methods}
\label{Discussion}
\begin{figure}
\centering
\includegraphics[width=0.4\textwidth]{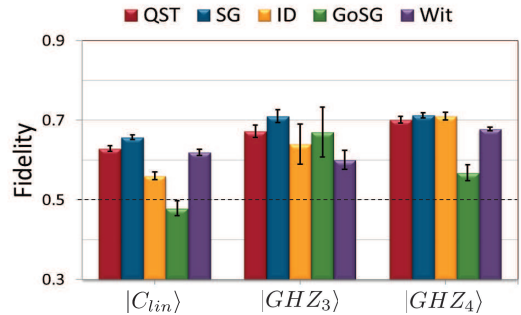}
	\caption{(Color online) 
	Comparison of fidelities obtained with different methods for the  four-qubit linear cluster, four-qubit GHZ state, and three-qubit GHZ state.
	The QST (red/first bar for every state) and SG (blue/second bar) approaches scale exponentially, while the ID (yellow/third bar), GoSG (green/fourth bar), and Wit (purple/fifth bar) approaches scale linearly with the number of qubits.
 Within the error bars the IDs set lower bounds, in agreement with the QST results. The SG fidelities tend to overestimate the QST ones. The GoSG and Wit bounds, like  the IDs, are  consistent with the rest of the methods. Note that $F_{GoSG}< 0.5$ for the four-qubit linear cluster,  so it is not sufficient to certify that the state can violate a Bell-type inequality.
The error bars derive from Poissonian statistics and thus correspond to a lower limit. }
\label{figure4}
\end{figure}

An interesting question is how IDs compare to other approaches
 used for state characterization of multiqubit states based on incomplete data. \\ 
Concerning the nonlocality proof,  we emphasize that the ID Bell inequality is composed of a minimal and irreducible set of mutually commuting observables for a specific state.
This is in contrast to previous works \cite{Scarani,Walther2005} where the joint observables
are not maximally entangled, implying that nonlocality could still be proven by preparing a state with fewer entangled qubits and using fewer parties. \new{ While our nonlocality test does not rule out hybrid hidden-variable models of entanglement or nonlocality \cite{Collins2002}, it does simultaneously discriminate against less entangled states within the Hilbert space formalism, as well as some different maximally entangled $N$-qubit states.} 
The Bell inequality for graph states proposed in Ref.~\cite{Guhne2005} involves  the complete stabilizer group (SG), which is always maximally entangled  but is not as compact as an ID, \new{ scaling exponentially with $N$ rather than linearly.} \\
Several witnesses were introduced to discriminate specific entangled states \cite{Bourennane2004,Tokunaga2006, Guehne2007}, providing analytic solutions,  which require  minimal experimental effort. Nevertheless, there was no generalization for the whole class of stabilizer states, only distinct derivations per subclass. For example,  Ref.~\cite{Guehne2007}  proposes a reduced witness for $N$-qubit cluster (GHZ) states which requires $N$ ($N+1$) measurement settings. The ID witness requires at most $N+1$ measurement settings for every stabilizer state, and for many specific cases it needs less than $N$ settings (e.g. the $ID5^4_w$ can be measured with four  settings and the $ID5^4_p$ with only three).  \new{ Each of these methods is minimal in some particular way, and both are robust against noise.
An additional method for entanglement discrimination,
 discussed in detail in \cite{Niekamp2010}, is to select subsets of stabilizer
 observables that are optimal for discriminating against
 a particular state, although a general method for obtaining these sets
 for $N$ qubits is lacking.  Unlike critical IDs, these subsets are usually
 not suitable as general entanglement witnesses, because
 they do not simultaneously discriminate against other particular
 states or less-entangled states.  Reference~\cite{Niekamp2010} also gives  a general method
 for discriminating between $N$-qubit stabilizer states using their
 complete stabilizer groups, but this method scales exponentially.  
The minimal ID witness sets can be simultaneously used to discriminate against particular states and, in some cases, also to achieve the optimal discrimination against particular states (as with the four-qubit GHZ state using the $\mathbf{ID}4^4_p$ in the \newT{Appendix B}).} \\
A fidelity estimation with incomplete data 
is obtained using the  SG of the state  \cite{Toth2005,Kiesel2005,note3}.
  This method,  based on $2^N$ measurement settings, still scales exponentially, just like the QST. \new{ Comparing the QST (from \cite{James}) and SG analyses for our experimental data in Fig.\ref{figure4} (first two bars), we see that the SG fidelity  
results in a higher value than 
   the QST fidelity for states with noise. The  
   QST approach  is considered to underestimate the real value of the fidelity \cite{Schwemmer}, 
 whereas    the SG approach,  based on the assumption of an $a$  $priori$ known ideal state,   
   might jeopardize the actual applicability of the characterized state if the resulting fidelity  overestimates the real value.} 
Alternatively, a lower bound of the fidelity can be found using the generators of the stabilizer group (GoSG) \cite{Wunderlich2009,Wunderlich2011}, the above-mentioned witnesses (Wit) \cite{note4}, 
 or the IDs.
 These techniques scale linearly and provide  thoroughly fair bounds for practical applications. Nevertheless the Wit's derivation is not general for  stabilizer states like  the ID and the GoSG approaches are. We analytically compare the last two methods in  \newT{Appendix A}, showing the IDs give  stronger (equally fair) bounds on the fidelity within an experimental environment.
We calculate the fidelity for  the experimentally generated stabilizer states using these estimations and summarize the result  in Fig.~\ref{figure4}.\\ 
We remark that the real value of the IDs approach is to capture all  the different quantum features of a state at one time.
We can exploit this generality to calculate the minimum fidelity required for an experimental demonstration of multiqubit nonlocality using IDs.
Simply setting $\langle \alpha \rangle_{exp} = \langle \alpha \rangle_{LHVT}=M-2  $ and  inverting  expression (\ref{FidBound}), we obtain  
$\langle|a|^2\rangle_{nonlocal} > 1/2$.
This verifies the already-proved limit of $50\%$ fidelity, which is  necessary for violation of any Bell-type inequality based on the GHZ theorem
\cite{Lu2014, Bell1997}. 
In most cases it is also the bound for discriminating less than maximally entangled states.
%

\section{Conclusion}
\label{sec:conclusion}
We have reported the characterization of an experimental four-qubit cluster state and a three-qubit GHZ state with the use of critical whole negative IDs.
Our efficient method requires only $N+1$ measurements for an $N$-qubit state and is of high practical value because it  provides simultaneously   a quantum state fidelity bound, an entanglement witness, and a nonlocality proof.
 For these reasons, IDs provide convenient laboratory tests of generated entangled resource states and certify that they are eligible for quantum science applications. Since the ID's observables belong to a single stabilizer group,  they can even be implemented
within stabilizer-based protocols such as quantum error correction and measurement-based quantum computing.\\
Entangled IDs, even if they are not critical, whole, or negative, can still be used to estimate the fidelity of a multiqubit state and to construct witness operators, as we have shown with the generated four-qubit GHZ state.
Additionally, special sets of IDs give rise to irreducible proofs of the
 $N$-qubit Kochen-Specker theorem \cite{Waegell2013,note5},
 demonstrating the conflict between non-contextual hidden variable theories and QM.  All of these connections emphasize the fundamental relationships between entanglement, contextuality, and nonlocality in quantum physics.\\
Furthermore, in the sense that these nonclassical phenomena are exactly the set of resources we wish to exploit, the full family of IDs is also the complete set of elemental resources for quantum information  processing within the $N$-qubit Pauli group.\\
\section{Acknowledgments}
We thank P.K. Aravind for several useful discussions.
This work was supported by the European Commission,
QUILMI (No. 295293), EQUAM (No. 323714), PICQUE (No. 608062), GRASP (No.
613024), QUCHIP (No. 641039), and the Vienna Center for Quantum Science and Technology (VCQ), the Austrian Science Fund (FWF) through
 START (No. Y585-N20),
and the doctoral program CoQuS, the Vienna Science and Technology Fund
(WWTF) under Grant No. ICT12-041, and the Air Force Office of Scientific
Research, Air Force Material Command, US Air Force, under Grant
No. FA8655-11-1-3004.

\section{Appendix A: Theory}
\label{sec:appendixA}
\subsection{Derivation of the ID Bell inequality}
In the following we show how to derive the ID Bell inequality given in Eq.(2) in the main text.\\
We rewrite the ID Bell parameter for a given negative  $\mathbf{ID}M^N_w$ as
\begin{equation}
\alpha = \sum_{i}^M \lambda_i O_i = \sum_{i|\lambda_i=1}^M  O_i - \sum_{i|\lambda_i=-1}^M O_i,
\label{alpha}
\end{equation}
where $O_i$ are the joint observables of the $\mathbf{ID}M^N_w$ and
$\lambda_i$ ($i=1,...,M$)
are the eigenvalues of the  ID eigenstate.
If a local hidden-variable theory (LHVT) is to agree with quantum mechanics (QM), then every term $O_i$ in $\alpha$ must be positive overall. This means that each
$O_i$ with $\lambda_i=1$ in Eq.(\ref{alpha})
must contain an even number of single-qubit Pauli observables $o_q$ assigned the value $-1$ and each $O_i$ with $\lambda_i=-1$
must contain an odd number of those.
Suppose that there are $n$ terms in $\sum_{i|\lambda_i=1}^M  O_i$
that contain two -1 values each, $m$ terms that  contain four -1 values each, $l$ terms with six, etc.  Likewise, there are $r$ terms in $\sum_{i|\lambda_i=-1}^M  O_i$
that  contain a single -1 value each, $s$ terms that contain three -1 values each, $t$ terms with five, etc.  We also note that because the ID is negative, the value $\gamma = r + s + t + \ldots$, which is the overall number of terms in the first summation, is always odd.
Using these definitions, we can write the total number of -1 values appearing in $\alpha$ as
\begin{align}
\eta &= (2n + 4m + 6l + \ldots) + (r + 3s + 5t + \ldots) \nonumber \\
& = (2n + 4m + 6l + \ldots) + (2s + 4t + \ldots) + \gamma.
\label{Bell Ineq Proof}
\end{align}
In the rightmost side of this equation, it is easy to see that the numbers in the parentheses are even, and then because $\gamma$ is odd, $\eta$ must also be odd.
Because the ID is whole, only even values of $\eta$ are possible in an LHVT, and this causes at least one term $O_i$ in $\alpha$ to be negative.  From this we obtain an upper bound, $\langle \alpha \rangle_{lhvt} \leq M-2$, which is finally our ID Bell inequality.
\subsection{Derivation of the ID fidelity bound}
For a general $\mathbf{ID}M^N_w$ (provided that it contains $M-1$ independent generators), we consider first the case that $|\psi\rangle$ is a pure state expressed in the eigenbasis of the ID,
\begin{equation}
|\psi\rangle = a|\kappa_0\rangle + \sum_{i=1}^V b_i |\kappa_i\rangle, 
\label{fid0}
\end{equation}
where $|\kappa_i\rangle$ are the $V-1$ eigenstates in the basis different from $|\kappa_0\rangle$ and $V$ is the number of all the possible states in the basis.  Of course $|a|^2 + \sum_{i=1}^V |b_i|^2 = 1$.
Then, the expectation value of $\alpha$
\begin{equation}
\langle \alpha \rangle_{exp} = |a|^2\langle\kappa_0|\alpha|\kappa_0\rangle + \sum_{i=1}^V |b_i|^2 \langle\kappa_i|\alpha|\kappa_i\rangle.
\label{fid1}
\end{equation}
We recall that the maximum value of $\langle \alpha \rangle_{QM}$ is $M$. Also, because the product of all eigenvalues is fixed for the observables of an ID, any eigenstate $|\kappa_i\rangle$ of the same ID with different values for $\lambda_i$ necessarily causes at least two terms in $\langle \alpha \rangle$ [Eq.\ref{alpha}] to be $-1$, resulting in a maximum of $M-4$ for that eigenstate.
If we allow the presence of noise, Eq.(\ref{fid1}) becomes
\begin{equation}
\langle \alpha \rangle_{exp} \leq M|a|^2 + \sum_{i=1}^V  (M-4)|b_i|^2 = 4|a|^2 + M-4,
\end{equation}
which we can rewrite as
\begin{equation}
|a|^2 \geq (\langle \alpha \rangle_{exp} - M + 4)/4.
\label{FidBound}
\end{equation}
This is the experimental lower bound on the probability amplitude of $|\kappa_0\rangle$ within the experimental state $|\psi\rangle$. It corresponds to a lower bound on the fidelity of a particular state for $M=N+1$ and the fidelity that the state lies within a particular subspace for $M<N+1$.\\
Next, we generalize the above derivation to the case of mixed states.  For a general convex combination of $m$ pure states plus noise,
\begin{equation}
\rho = c_0 \frac{I}{2^N} + \sum_{j=1}^m c_j |\psi_j\rangle\langle\psi_j|,
\end{equation}
where $\sum c_j = 1$, we can expand each $|\psi_j\rangle$ as in Eq.(\ref{fid0}) , $|\psi_j\rangle = a_j|\kappa_0\rangle + \sum_{i=1}^V b_{ij} |\kappa_{i}\rangle$, and follow the same argument to obtain
\begin{equation}
\langle \alpha_{exp} \rangle \leq \sum_{j=1}^m c_j (4|a_j|^2+M-4).
\end{equation}
Given that we have no experimental access to  $c_j$, we must allow the constant term to take its maximum value, and then we obtain
\begin{equation}
\langle|a|^2\rangle \geq (\langle \alpha_{exp} \rangle - M+4)/4,
\end{equation}
where $\langle|a|^2\rangle \equiv \sum c_j |a_j|^2$ is the weighted average amplitude of $|\kappa_0\rangle$ among the pure states that make up $\rho$ and the noise component (for which the amplitude of $|\kappa_0\rangle$ is assumed to be $a_0 = 0$).  Therefore the most general interpretation of our inequality is that it places a lower bound on the average amplitude of $|\kappa_0\rangle$ within a mixed state $\rho$ and thus that we have obtained a lower bound on the fidelity of the prepared state.  This also allows for the possibility that our $N$ qubits are entangled with additional ancillary qubits that we do not control, since measuring them is then analogous to measuring some convex mixture of $N$-qubit pure states.
\subsection{Comparing the fidelity bounds obtained using IDs and generators}
Let us now compare the fidelity bounds obtained with our ID-based method and the generator-based method (GoSG) of Ref.~\cite{Wunderlich2009}.  In that work the authors provide a general equation for any set of $N$ generators which gives the fidelity to be bounded below by
\begin{equation}
F_{GoSG} = (\sum_n a_n - N + 2)/2,
\end{equation}
while our ID-based method gives a lower bound of
\begin{equation}
F_{ID} = (\sum_m a_m - M + 4)/4,
\end{equation}
where $A_i$ are observables and $a_i = \langle A_i \rangle$ are their experimentally obtained expectation values.\\
Their method makes use of the $N$ specific generators of a graph state, for which all eigenvalues $\lambda_n = 1$.  Every set of $N$ independent generators gives an ID$M^N$ (with $M=N+1$) by adding one more observable $A_M$ to the set,
\begin{equation}
A_M = \lambda_M \prod_n A_n,
\end{equation}
with $\lambda_M$ being equal to the sign of the resulting ID, such that $0 \leq a_i \leq 1$.

Putting all of this together we can construct a quantitative comparison of our two bounds for the same set of $N$ generators and the $M$th observable  needed for our method.
\begin{align}
F_{ID} - F_{GoSG} &= (\sum_m a_m - M + 4)/4 - (\sum_n a_n - N + 2)/2 \nonumber \\
& = [(a_M-1) + (N - \sum_n a_n)]/4.
\label{eq_ID_Gen}
\end{align}
Clearly the difference vanishes when both methods give perfect fidelity.  However, in the case that the measurements are imperfect, $-1 \leq a_M - 1 \leq 0$ and $0 \leq N-\sum_n a_n \leq N$.  If we let all of the $a_m$ take the same average value (call it $a_0 < 1$), then this reduces to
\begin{equation}
F_{ID} - F_{GoSG} = (N-1)(1-a_0)/4 > 0,
\end{equation}
which shows that our bound is usually better.  Of course in practice this will depend on the specific values of  $a_m$, and indeed, in the bizarre case where  $a_M = 0$ and $a_n = 1$, we get $F_{ID}=0.75$ and $F_g=1$, and their bound is actually better by $1/4$.  So, generally speaking, the best practice will be to take the better of these two bounds for a given set of measured values $a_m$, and their method gives a better bound when
\begin{equation}
N - \sum_n a_n < 1 - a_M
\end{equation}
or
\begin{equation}
\sum_n e_n < e_M,
\end{equation}
where $e_i = 1-a_i$ is the error of each measurement.  Interestingly, it is truly arbitrary which of the observables $A_m$ in an ID is chosen to be  $A_M$, which means we can examine all $M$ choice, and take the best of the $M+1$ different bounds obtained from the measured set $a_m$.  $F_{ID}$ is better for the case when the average errors of the all measurements are  comparable, but if the error of any one measurement is worse than all the others combined, then $F_{GoSG}$ is the superior bound, effectively allowing us to ignore the one particularly bad measurement.  The relative quality of the good and bad measurements required to satisfy this condition increases linearly with $N$, and thus it becomes increasingly unlikely that we can throw away a measurement in this way.  Therefore in a realistic experimental setting, as $N$ increases, $F_{ID}$ quickly becomes the superior bound.
\subsection{Derivation of the ID entanglement witness}
Here we give the derivation of the analytic solution for the upper bound $\Gamma_\mathcal{C}$ on $\gamma_\mathcal{C}$ for ID witness observables.
We begin by rewriting Eq. (\ref{FidBound}) as
\begin{equation}
\langle \alpha \rangle_{exp} \leq 4|\langle\kappa_0|\psi\rangle|^2 + M-4,
\end{equation}
where $|\kappa_0\rangle$ is the particular eigenstate whose eigenvalues are used to define $\alpha$ for this ID.
Next, we let $\mathcal{C}$ be the class of all  possible bipartitions $\{B_l\}$ of the $N$-qubit system.
Following the derivation in \cite{Bourennane2004}, we obtain
\begin{equation}
\max_{|\psi\rangle \in B_l} \langle \alpha \rangle_{exp}   \leq   M-4 + 4[\max_{m}\{ \nu_m \}]^2 \equiv \beta_l,
\end{equation}
where $\{ \nu_m \}$ are the Schmidt coefficients of $|\kappa_0\rangle$ with respect to the bipartition $B_l$.  We therefore find that $\Gamma_\mathcal{C} = \max_{l} \beta_l$.  In many cases the individual $\beta_l$ have values lower than $\Gamma_\mathcal{C}$,  so this method can be used to discriminate more strongly against some bipartitions $B_l$ than others.  There is also a more general analytic solution for $\gamma_\mathcal{C}$ that rules out some other nonbiseparable types of states with different Schmidt numbers.

As in other cases \cite{Tokunaga2006,Guehne2007}, we can also obtain a relation between these analytic entanglement witnesses and our measure of fidelity of the quantum state:
\begin{equation}
F_{ID} = (\gamma_\mathcal{C} - \langle\mathcal{W}^{ID}_\mathcal{C} \rangle - M + 4)/4.
\end{equation}

\new{When $|\psi\rangle$ is another stabilizer state, an upper bound can also be determined analytically, as shown in \cite{Niekamp2010}.  For our purposes, this method works by considering which observables from the ID and the state's stabilizer act nontrivially on the same qubits.  For the $\mathbf{ID}5^4$ cases presented here, $\Gamma_\mathcal{C}$ gives a bound equal to or better than that of this method;  the only exception is the case of using  $\mathbf{ID}5^4_p$ (related to the four-qubit GHZ state) to discriminate against the \text{shear} cluster  $|C_{\utimes}\rangle$ (where it gives $\gamma =2$ while $\Gamma_\mathcal{C}=3$, and the numerical result $\gamma=1$ is still better).  For the $\mathbf{ID}4^4_p$, $\Gamma_\mathcal{C}=4$ is useless because that method maximizes over two terms in a sum independently, ignoring their mutual constraints.  In this case, the method of \cite{Niekamp2010} can still be applied to analytically obtain the $\gamma_\mathcal{C}=0$ results in Table \ref{ID4Fig} (see Appendix B), but for all of the other cases in that table, it gives $\gamma=4$, and the numerical results are still better.  This is partly because their general method is tailored to discriminating between graph states with connected graphs and neglects less entangled states.}

As indicated above, in many cases we can obtain better values for $\gamma_\mathcal{C}$ by directly maximizing over $\langle \psi|\alpha|\psi\rangle$ numerically.  Obviously, no general solution is known for all possible classes of states $\mathcal{C}$, but numerical techniques can be used to obtain maxima for many particular cases, allowing us to discriminate against them, sometimes quite strongly.\\

We should point out that these witness techniques implicitly assume the Hilbert space formalism of quantum mechanics.  A more general type of witness can be constructed that rules out any hidden-variable theory without pairwise correlations between every pair of qubits in the state \cite{Collins2002}.  Such witnesses require one to measure a set of observables that do not all mutually commute,  so we cannot obtain this result within any stabilizer-based protocol.

\subsection{Noise tolerance of ID entanglement witness}
As has been done in the  other cases \cite{Tokunaga2006}, we can compute the general tolerance of our ID witness observables to white noise.  To compute the tolerance, we solve $\mathrm{Tr}(\mathcal{W}^{ID} \rho(p_N)) < 0$ for $p_N$, where $\rho(p_N) = p_N/2^N I + (1-p_N)|\psi\rangle \langle\psi|$ is the standard depolarizing noise channel and $|\psi\rangle$ is the state we intend to witness.  For an \textbf{ID}$M^N$ with $M=N+1$,
\begin{equation}
p_N < \frac{M-\gamma}{M}.
\end{equation}
More generally, the eigenbasis of an \textbf{ID}$M^N$ is composed of projectors $|\psi\rangle\langle\psi|$ of rank $r=2^{N-M+1}$, and the noise tolerance is given by
\begin{equation}
p_N < \frac{r(M-\gamma)}{r(M-\gamma) + \gamma}.
\end{equation}
These tolerances are valid regardless of what method is used to obtain $\gamma$.

\subsection{Entanglement in the Heisenberg picture}
 For $N \geq 4 $, there exist maximally entangled IDs with fewer than $N$ independent generators that lie at the intersection of the stabilizer groups of multiple locally inequivalent classes of entangled states.  Therefore we do not find a one-to-one correspondence between the classification of locally inequivalent entangled graph states (Schr\"{o}dinger picture) and the classification of entangled locally-inequivalent IDs (Heisenberg picture).  This mismatch leads to the existence of maximally entangled subspaces (belonging to IDs) that can contain a continuum of locally inequivalent states (including several locally inequivalent graph states).  Indeed, the code spaces already employed in quantum error correction are of exactly this type, although the general utility of maximally entangled spaces is more subtle and interesting.

 To get a sense of the structure that emerges here, we can look at the four- and five-qubit cases.  For four qubits, there are three  locally inequivalent cluster states (as discussed in the main paper); nevertheless, there exists a critical ID$4^4$ with an eigenbasis of rank-2 subspaces that contain all three types of cluster states.

 For five qubits, there are four  classes of maximally entangled stabilizer states up to local unitaries and reordering of qubits.  These are the five-qubit GHZ state, cluster state, pentagon state, and one other that we will call the cluster-B state.

 The GHZ stabilizers do not contain any ID$5^5$'s,  so the five-qubit GHZ-type entanglement does not belong to any maximally entangled subspaces of IDs.  The pentagon and cluster state share a common negative ID$5^5_w$, and thus there is a maximally entangled two-dimensional subspace that contains both of these types of states, and all states in this space provide proof of the GHZ theorem.  There are also critical ID$5^5$'s that are common to the cluster and cluster-B states, but none of these are whole and negative; thus while they do define maximally entangled spaces, they do not provide proof of the GHZ theorem.  There are also numerous spaces that span locally inequivalent versions (permutations of qubits) of a given entangled state,  just as in the four-qubit cluster case.

 From the above, we can see that the Bell and GHZ states look more or less the same in both the Heisenberg and Schr\"{o}dinger pictures, but the same is not true for the other types of states.  The other types are simply cardinal states within complete maximally entangled subspaces that remain intact under local unitary evolution.

\section{Appendix B: Analysis}
\label{sec:appendixB}
\subsection{Four-qubit linear cluster state}
\noindent \textbf{ID entanglement witness}\\
We present here the method we used to obtain numerical bounds for $\gamma_\mathcal{C}$ for the $\mathbf{ID}5^4_w$ in order to discriminate against states other than $|C_{lin}\rangle$. \\
We break the analysis into pieces based on each LU-inequivalent class of  an $N$-qubit state.  This significantly reduces the number of parameters needed to explore the general state space.  For four qubits, a general pure state has 30 free parameters.  If we begin with a particular entangled state, then we can explore the entire entanglement class using only LU operations, and this reduces the number of free parameters to at most 12 (which is a significant reduction in terms of computational resources needed to calculate the bounds).  We use the MATLAB OPTIMIZATION TOOLBOX  function FMINSEARCH.M to perform the multivariate maximization.  This function finds local maxima based on an initial guess.  We therefore proceed with a sort of $ad$ $hoc$ ``Monte Carlo'' maximization technique by making a large number of initial guesses and taking the best local maximum from among these runs.  In order to get convergent results from this method, we actually compute an upper bound $\gamma_\mathcal{C} \leq \max_\mathcal{C} \sum_{i}^M | \langle O_i \rangle |$.  This function has far fewer local maxima in $\mathcal{C}$ than $\langle\alpha\rangle$.
We report in Table \ref{gammaSI} the obtained upper bounds of $\gamma_\mathcal{C}$ for several quantum states.
  We considered a fully separable state, $|\psi_1\rangle |\psi_2\rangle |\psi_3\rangle |\psi_4\rangle$, product states  of two-qubit Bell states $|\Phi_{ij}\rangle$ ($i,j=\{1,2,3,4\}$), partial separable states, GHZ states $|GHZ\rangle$, W states $|W\rangle$, and different types of cluster states, $|C_{\utimes}\rangle=(\ket{0000}+\ket{0101} +\ket{1010}-\ket{1111})/2$ 			and $|C_{\text{\FallingEdge}}\rangle=(\ket{0000}+\ket{0110} +\ket{1001}-\ket{1111})/2$. \\
 Although we lack general numerical results for $N\geq4$, we conjecture that the the negativity of $\mathcal{W}^{ID}$ ( which  is equivalent to the violation of the ID Bell inequality) can happen only with the specific stabilizer state that corresponds to $\alpha$ (up to LU transformations) - or states that include it as a large enough part of a superposition and/or mixed state.\\

Within the cluster stabilizer group there are 196 different entangled IDs, belonging to 8 specific isomorphism classes  with $M=5$ or $M=4$  and distinct features. From each of these we can obtain an ID fidelity and an ID entanglement witness. As an example we show in Table \ref{ID44} one such positive partial $\mathbf{ID}4^4_p$.
The corresponding ID-witness allows us to discriminate much more strongly against some entangled states with the  numerical maximization method than with the analytic solution for the same witness [see Tables \ref{ID44_gamma} and \ref{ID44_gamma1}].  Of particular interest are the cases where $\gamma_\mathcal{C} = 0$ since we can discriminate against these states with perfect noise tolerance:  any $\langle\alpha\rangle_{exp} >0$ is sufficient.

\begin{table}[H]
\centering
\subfloat{
\begin{tabular}{l|c}
State type & $\gamma_\mathcal{C}$\\
\hline
\rule{0pt}{10pt}$|\psi_1\rangle |\psi_2\rangle |\psi_3\rangle |\psi_4\rangle^*$ & $2$\\
$|\psi_1\rangle |\psi_2\rangle |\Phi_{34}\rangle $ & $3$\\
$|\psi_1\rangle |\psi_3\rangle |\Phi_{24}\rangle^* $ & $2$\\
$|\psi_1\rangle |\psi_4\rangle |\Phi_{23}\rangle^* $ & $2$\\
$|\psi_2\rangle |\psi_3\rangle |\Phi_{14}\rangle^* $ & $2$\\
$|\psi_2\rangle |\psi_4\rangle |\Phi_{13}\rangle^* $ & $2$\\
$|\psi_3\rangle |\psi_4\rangle |\Phi_{12}\rangle $ & $2$\\
$|\Phi_{12}\rangle |\Phi_{34}\rangle $ & $3$\\
$|\Phi_{13}\rangle |\Phi_{24}\rangle^* $ & $1$\\
$|\Phi_{14}\rangle |\Phi_{23}\rangle^* $ & $1$\\
$|\psi_{1}\rangle |GHZ_{234}\rangle $ & $3$\\
$|\psi_{2}\rangle |GHZ_{134}\rangle $ & $3$\\
\end{tabular}
}
\qquad
\subfloat{
\begin{tabular}{l|c}
State type & $\gamma_\mathcal{C}$\\
\hline
\rule{0pt}{10pt}$|\psi_{3}\rangle |GHZ_{124}\rangle $ & $3$\\
$|\psi_{4}\rangle |GHZ_{123}\rangle $ & $3$\\
$|\psi_{1}\rangle |W_{234}\rangle $ & $2.6667$\\
$|\psi_{2}\rangle |W_{134}\rangle $ & $2.6667$\\
$|\psi_{3}\rangle |W_{124}\rangle $ & $2.3610$\\
$|\psi_{4}\rangle |W_{123}\rangle $ & $2.3610$\\
$|GHZ_{1234}\rangle $ & $3$\\
$|W_{1234}\rangle $ & $3$\\
$|C_{\utimes}\rangle $ & $3$\\
$|C_{\text{\FallingEdge}}\rangle $ & $3$\\
$|C_{lin}\rangle $ & $\textbf{5}$\\
$ $& $ $\\
\end{tabular}}
\label{gammaSI}
\captionsetup{justification=raggedright}
\caption{ Numerical upper bounds on $\gamma_\mathcal{C}$ ($max_{|\psi\rangle \in \mathcal{C}} \langle\alpha\rangle $) for $\mathbf{ID}5^4_w$.
All the  quantum states, which differ from the target state $|C_{lin}\rangle $ , have the analytic bound  $\Gamma_\mathcal{C} = 3$, except for particular bipartitions (marked with an asterix) where $\Gamma_\mathcal{C} = 2$. In some cases the numerical values result are even lower.}
\end{table}

\begin{table}[H]
\hspace{-0.1cm}
\subfloat[][]{
\begin{tabular}{cccc}
$Z$ & $Z$ & $Z$ & $I$\\
$X$ & $X$ & $I$ & $Z$\\
$Y$ & $I$ & $X$ & $X$\\
$I$ & $Y$ & $Y$ & $Y$\\
\end{tabular} \label{ID44}}
\hspace{0.2cm}
\subfloat[][]{ 
\begin{tabular}{l|c}
State type & $\gamma_\mathcal{C}$\\
\hline
\rule{0pt}{10pt}$|\psi_1\rangle |\psi_2\rangle |\psi_3\rangle |\psi_4\rangle$ & $1$\\
$|\psi_1\rangle |\psi_2\rangle |\Phi_{34}\rangle $ & $2$\\
$|\psi_1\rangle |\psi_3\rangle |\Phi_{24}\rangle $ & $2$\\
$|\psi_1\rangle |\psi_4\rangle |\Phi_{23}\rangle$ & $2$\\
$|\psi_2\rangle |\psi_3\rangle |\Phi_{14}\rangle$ & $2$\\
$|\psi_2\rangle |\psi_4\rangle |\Phi_{13}\rangle$ & $2$\\
$|\psi_3\rangle |\psi_4\rangle |\Phi_{12}\rangle$ & $2$\\
$|\Phi_{12}\rangle |\Phi_{34}\rangle $ & $0$\\
$|\Phi_{13}\rangle |\Phi_{24}\rangle $ & $0$\\
$|\Phi_{14}\rangle |\Phi_{23}\rangle $ & $0$\\
$|\psi_{1}\rangle |GHZ_{234}\rangle $ & $2$\\
$|\psi_{2}\rangle |GHZ_{134}\rangle $ & $2$\\
\end{tabular}
\hspace{0.6cm}
\begin{tabular}{l|c}
State type & $\gamma_\mathcal{C}$\\
\hline
\rule{0pt}{10pt}$|\psi_{3}\rangle |GHZ_{124}\rangle $ & $2$\\
$|\psi_{4}\rangle |GHZ_{123}\rangle $ & $2$\\
$|\psi_{1}\rangle |W_{234}\rangle $ & $2$\\
$|\psi_{2}\rangle |W_{134}\rangle $ & $2$\\
$|\psi_{3}\rangle |W_{124}\rangle $ & $2$\\
$|\psi_{4}\rangle |W_{123}\rangle $ & $2$\\
$|GHZ_{1234}\rangle $ & $0$\\
$|W_{1234}\rangle $ & $2$\\
$|C_{\utimes}\rangle $ & $4$\\
$|C_{\text{\FallingEdge}}\rangle $ & $4$\\
$|C_{lin}\rangle $ & $\textbf{4}$\\
$ $& $ $\\
\end{tabular} }
\captionsetup{justification=raggedright}
\caption{  (a) Representation of $\mathbf{ID}4^4_p$ . (b) Numerical upper bounds on $\gamma_\mathcal{C}$ for $\mathbf{ID}4^4_p$. \new{ The analytic method always fails for this case (i.e.,  $\Gamma_\mathcal{C}=4$). 
The method proposed in  Ref.~\cite{Niekamp2010} gives all cases with $\gamma_\mathcal{C}=0$, but fails (i.e. $\gamma=4$) for all other cases.} 
}\label{ID4Fig}
\end{table}

\begin{figure}[h]
\includegraphics[scale=0.35]{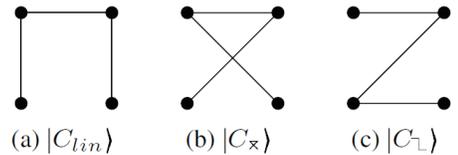}
\caption{Different types of cluster states.}
\label{cluster2}
\end{figure}

We show the graphs that generate each of the three LU-inequivalent four-qubit cluster states in Fig.~\ref{cluster2}. The graphs in Fig.~\ref{cluster2}(b) and (c) are obtained by exchanging the order of qubits in the linear cluster state $\ket{C_{lin}}=(\ket{0000}+\ket{0011}+\ket{1100}-\ket{1111})/2$.

\begin{figure}[H]
	\centering
	\includegraphics[width=0.4\textwidth]{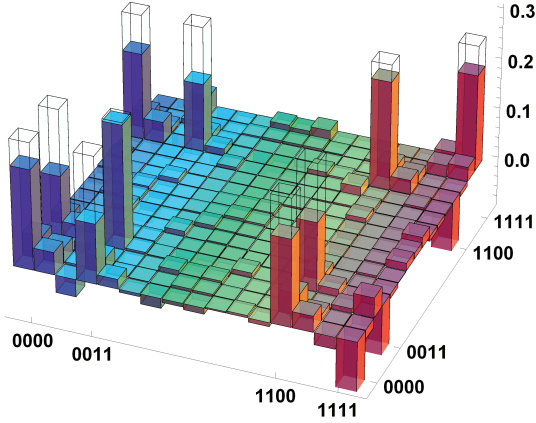}
	\caption{(Color online) Reconstructed density matrix (real part) of the four-qubit cluster state ($F_{QST}=0.629 \pm 0.007$). The imaginary part is not shown since its components are below $0.05$.}
\label{Figure1_SI}
\end{figure}

\noindent \textbf{Quantum state tomography}\\
We reconstruct the density matrix of the generated cluster state through complete quantum state tomography. The real part is shown in Fig.~\ref{Figure1_SI}. The components of the imaginary part are below $0.047$ and are hence not presented here. \\
The error is estimated running a 100-cycle Monte Carlo simulation with Poissonian noise added to the experimental counts.

\noindent \textbf{Stabilizer group}\\
The stabilizer group operators and their respective expectation values are reported  in Table~\ref{figure2SI}.\\
\begin{table}[H]
\subfloat{
\begin{tabular}{l|r }
\toprule
Observable & Expectation value \\ 
\hline
$Z Z I I$ &	$0.93  \pm  0.01$\\
$I I Z Z$ &  $0.78  \pm	0.02$\\
$Z I X X$ &	$0.61  \pm  0.02$\\
$I Z X X$ &	$0.59  \pm  0.02$\\
$I Z Y Y$ &	$-0.58 \pm	0.02$\\
$Z I Y Y$ &  $-0.58 \pm  0.02$\\
$X X Z I$ &  $0.66  \pm	0.02$\\
$X X I Z$ &  $0.62  \pm	0.02$\\ \bottomrule
\end{tabular} }
\hspace{0.35cm}
\subfloat{
\begin{tabular}{l|r }
\toprule
Observable & Expectation value \\ 
\hline
$Y Y I Z$ &  $-0.65 \pm	0.02$\\
$Y Y Z I$ &	$-0.65 \pm  0.02$\\
$X Y X Y$ &	$0.47  \pm  0.02$\\
$X Y Y X$ &	$0.52  \pm  0.02$\\
$Y X X Y$ &	$0.52  \pm	0.02$\\
$Y X Y X$ &	$0.60  \pm	0.02$\\
$Z Z Z Z$ &	$0.75  \pm  0.02$\\
$I I I I$ &	$1	  \pm  0.03$\\ \bottomrule
\end{tabular}}\caption{Measured expectation values for all operators in the stabilizer group of $\ket{C_{lin}}$. For $F_{GoSG}$ we used the operators $ZZII$, $IIZZ$, $IZXX$, $XXZI$.}
\label{figure2SI}
\end{table}

\begin{figure}[H]
\centering
\includegraphics[scale=0.65]{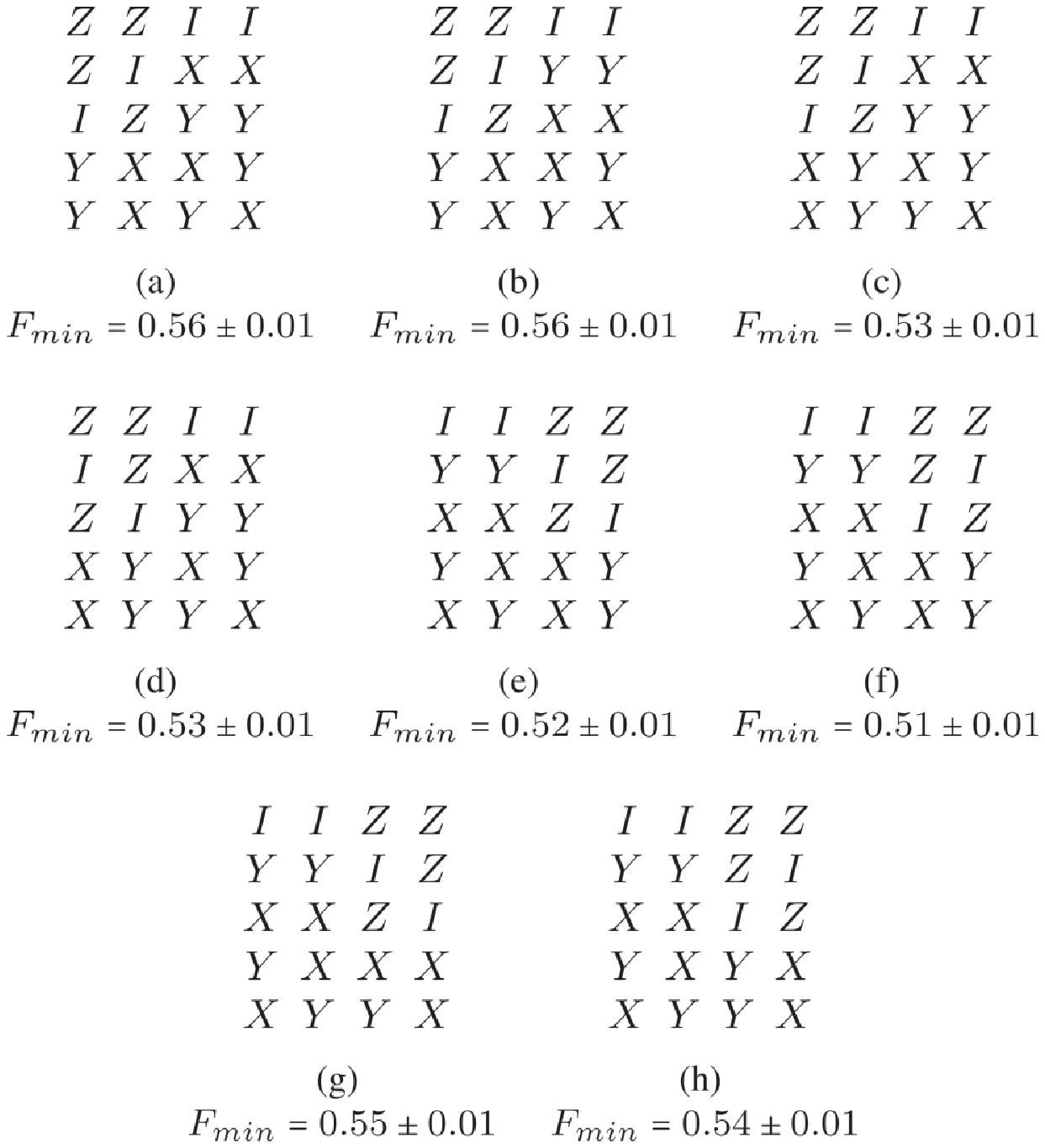}
\caption{All eight  equivalent ID$5^{4}_{w}$  whose joint eigenstate is $\ket{C_{lin}}$.}
\label{equivID}
\end{figure}

\noindent \textbf{Equivalent IDs}\\
We show in Fig.~\ref{equivID} the eight  equivalent $\mathbf{ID}5^4_w$'s within the stabilizer group of $\ket{C_{lin}}$.  We calculate the relative bounds of fidelity for each of these IDs, obtaining results in the range $\{0.51\pm 0.01, 0.56\pm 0.01\}$.

\subsection{Three-qubit GHZ state}
\noindent \textbf{Stabilizer group}\\
The stabilizer group operators and their respective expectation values are reported  in Table~\ref{figureSG}. Note that these results are extrapolated from the quantum state tomography setting of the cluster state and after  projection of the second qubit of the cluster state onto the state $|-\rangle=(|0\rangle -|1\rangle)/ \sqrt{2}$.\\

\begin{table}[H]
\centering
\setlength{\tabcolsep}{0.3cm}
\begin{tabular}{l|r }
\toprule
	Observable & Expectation value \\ 
	\hline
$X X X$ &	$0.81  \pm  0.07$\\
$Y X Y$ &	$-0.61 \pm	0.09$\\
$X Y Y$ &  $-0.59 \pm  0.09$\\
$Y Y X$ &	$ -0.54  \pm  0.10$\\
$Z Z I$ &  $0.61  \pm	0.09$\\
$Z I Z$ &  $-0.64  \pm	0.09$\\
$I Z Z$ &  $0.88 \pm	0.05$\\
$I I I$ &	$1 \pm  0.12$\\ \bottomrule
\end{tabular}\caption{Measured expectation values for all operators in the stabilizer group of $\ket{GHZ_{3}}$. The first four values are used to obtain a $F_{ID}= 0.64\pm 0.04$. $ZZI$,$IZZ$, and $XXX$ are the generators used for $F_{GoSG}$.}
\label{figureSG}
\end{table}

\begin{figure}[H]
	\centering
	\includegraphics[width=0.35\textwidth]{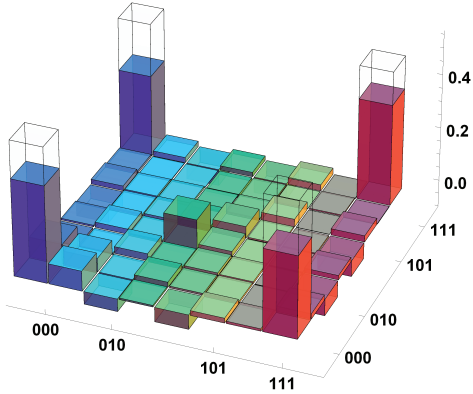}
	\caption{(Color online) Reconstructed density matrix (real part) of the three-qubit GHZ state ($F_{QST}=0.672 \pm 0.015$). The imaginary part has components below $0.07$ and is not shown.}\label{Figure2_SI}
\end{figure}

\vspace{2cm}
\noindent \textbf{Quantum state tomography}
We present in Fig.~\ref{Figure2_SI} the density matrix of the experimental three-qubit GHZ state, reconstructed through complete quantum state tomography.

\subsection{Four-qubit GHZ state}
\noindent \textbf{ID entanglement witness}\\
We report in Table \ref{GHZ4_gamma} the numerical values of $\gamma_\mathcal{C}$ for the $\mathbf{ID}5^4_p$ calculated via the same maximization procedure used for the four-qubit cluster case. The analytic bound is $\gamma_\mathcal{C} = 3$ for all bipartitions.\\

\begin{table}[H]
\centering
\subfloat{
\begin{tabular}{l|c}
State type & $\gamma_\mathcal{C}$\\
\hline
\rule{0pt}{10pt}$|\psi_1\rangle |\psi_2\rangle |\psi_3\rangle |\psi_4\rangle$ & $3$\\
$|\psi_1\rangle |\psi_2\rangle |\Phi_{34}\rangle $ & $2$\\
$|\psi_1\rangle |\psi_3\rangle |\Phi_{24}\rangle $ & $2$\\
$|\psi_1\rangle |\psi_4\rangle |\Phi_{23}\rangle $ & $1$\\
$|\psi_2\rangle |\psi_3\rangle |\Phi_{14}\rangle $ & $1$\\
$|\psi_2\rangle |\psi_4\rangle |\Phi_{13}\rangle $ & $2$\\
$|\psi_3\rangle |\psi_4\rangle |\Phi_{12}\rangle $ & $2$\\
$|\Phi_{12}\rangle |\Phi_{34}\rangle $ & $3$\\
$|\Phi_{13}\rangle |\Phi_{24}\rangle $ & $3$\\
$|\Phi_{14}\rangle |\Phi_{23}\rangle $ & $1$\\
$|\psi_{1}\rangle |GHZ_{234}\rangle $ & $3$\\
$|\psi_{2}\rangle |GHZ_{134}\rangle $ & $3$\\
\end{tabular}}
\qquad
\subfloat{
\begin{tabular}{l|c}
State type & $\gamma_\mathcal{C}$\\
\hline
\rule{0pt}{10pt}$|\psi_{3}\rangle |GHZ_{124}\rangle $ & $3$\\
$|\psi_{4}\rangle |GHZ_{123}\rangle $ & $3$\\
$|\psi_{1}\rangle |W_{234}\rangle $ & $2.3333$\\
$|\psi_{2}\rangle |W_{134}\rangle $ & $2.3333$\\
$|\psi_{3}\rangle |W_{124}\rangle $ & $2.3333$\\
$|\psi_{4}\rangle |W_{123}\rangle $ & $2.3333$\\
$|W_{1234}\rangle $ & $3$\\
$|C_{lin}\rangle $ & $3$\\
$|C_{\utimes}\rangle $ & $3$\\
$|C_{\text{\FallingEdge}}\rangle $ & $1$\\
$|GHZ_{1234}\rangle $ & $\textbf{5}$\\
$ $& $ $\\
\end{tabular}}\captionsetup{justification=raggedright}
\caption{ Numerical upper bounds on $\gamma_\mathcal{C}$ for $\mathbf{ID}5^4_p$.  For biseparable states, the analytic bound is $\Gamma_\mathcal{C} = 3$, while in some cases the numerical result is lower.\\}
\label{GHZ4_gamma}
\end{table}

\begin{figure}[H]
	\centering
	\includegraphics[width=0.4\textwidth]{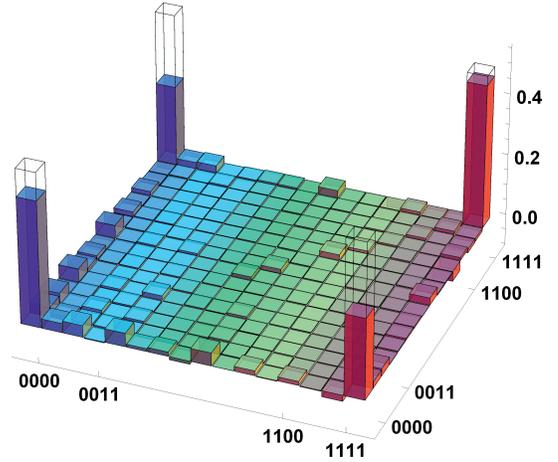}
	\caption{(Color online) Reconstructed density matrix (real part) of the four-qubit GHZ state ($F_{QST}=0.701 \pm 0.008$). The imaginary part is not shown since its components are below $0.03$.\\}\label{Figure3}
\end{figure}

\vspace{5cm}
 \noindent \textbf{Quantum state tomography}
 We present in Fig.~\ref{Figure3} the density matrix of the experimental four-qubit GHZ state, reconstructed through complete quantum state tomography.

\noindent \textbf{Stabilizer group}\\
The stabilizer group operators and their respective expectation values are reported  in Table~\ref{results4qGHZ_stab}.

\begin{table}[H]
\subfloat[][]{
\begin{tabular}{l|r}\toprule
\textbf{Stabilizer} & \textbf{Expectation value }\\
\hline
$ZZII$	&$0.87	 \pm 0.02$\\
$IIZZ$	&$0.88	\pm 0.02$\\
$ZIZI$	&$0.90	\pm 0.02$\\
$IZIZ$	&$0.90	\pm 0.02$\\
$ZIIZ$	&$0.85	\pm 0.02$\\
$IZZI$	&$0.85	\pm 0.02$\\
$ZZZZ$	&$0.85	\pm 0.02$\\
$XXXX$	&$0.54	\pm 0.03$\\ \bottomrule
\end{tabular} }
\hspace{0.3cm}
\subfloat[][]{
\begin{tabular}{l|r }
\toprule
	Observable & Expectation value \\ 
	\hline
$YYYY$	&$0.56	\pm 0.03$\\
$XXYY$	&$-0.51	\pm 0.03$\\
$XYXY$	&$-0.56	\pm 0.03$\\
$XYYX$	&$-0.60	\pm 0.03$\\
$YXXY$	&$-0.48	\pm 0.03$\\
$YXYX$	&$-0.51  \pm 	0.03$\\
$YYXX$	&$-0.53	\pm 0.03$\\
$IIII$	&$1	\pm 0.03$\\
\bottomrule
\end{tabular}}
\caption{Measured expectation values for the observables in the stabilizer group of $\ket{GHZ_{4}}$. The acquisition time for each measurement setting was 4800 s.}
\label{results4qGHZ_stab}
\end{table}

\bibliographystyle{apsrev}
\bibliography{_mybib4}

\begin{thebibliography}{42}
\bibitem[{not({\natexlab{a}})}]{note1}
\bibinfo{note}{The product of all the joint observables is equal to $-1$
  ($\mathbf{ID}=-\mathbf{I}$). LHVTs require that a truth-value is preassigned
  to all single-qubit observables. For LHVTs the overall product of the
  eigenvalues of each single-qubit observable is always $+1$, since each
  single-qubit observable appears twice in the ID. This brings us to the famous
  GHZ contradiction $-1=+1$.}
\bibitem[{not({\natexlab{b}})}]{note2}
\bibinfo{note}{Note that $\langle\mathcal{W}^{ID}_\mathcal{C}
  \rangle=\gamma_\mathcal{C}I - \langle\alpha_{exp} \rangle$, where $\langle
  \alpha_{exp} \rangle$ is the measured value of the ID-Bell paramter.}
\bibitem[{not({\natexlab{c}})}]{note6}
\bibinfo{note}{Here maximally entangled means that no subset of qubits is
  separable from another subset.}
\bibitem[{not({\natexlab{d}})}]{note3}
\bibinfo{note}{The stabilizer group is itself always an ID, although it can be
  critical only for $N=2$ and can be otherwise factored into products of
  smaller IDs. Each critical ID of a given $N$ belongs to one or more
  stabilizer groups and within these abelian subgroups, the ID can be seen as a
  sort of ''prime'' structure, in the sense that it cannot be factored into
  smaller structures in smaller stabilizer groups.}
\bibitem[{not({\natexlab{f}})}]{note5}
\bibinfo{note}{Any maximally entangled whole negative ID with $M\leq N+1$ gives
  an irreducible proof of the GHZ theorem, while full criticality of IDs is the
  crucial feature for irreducible proof of the KS theorem.}
\end{thebibliography}
\end{document}